\documentclass[conference]{IEEEtran}
\IEEEoverridecommandlockouts
\usepackage{cite}
\usepackage{amsmath,amssymb,amsfonts}
\usepackage{algorithmic}
\usepackage{graphicx}
\usepackage{multirow}
\usepackage{textcomp}
\usepackage[normalem]{ulem}
\usepackage{xcolor}
\def\BibTeX{{\rm B\kern-.05em{\sc i\kern-.025em b}\kern-.08em
    T\kern-.1667em\lower.7ex\hbox{E}\kern-.125emX}}

\usepackage{array}
\newcolumntype{L}[1]{>{\raggedright\let\newline\\\arraybackslash\hspace{0pt}}m{#1}}
\newcolumntype{C}[1]{>{\centering\let\newline\\\arraybackslash\hspace{0pt}}m{#1}}
\newcolumntype{R}[1]{>{\raggedleft\let\newline\\\arraybackslash\hspace{0pt}}m{#1}}

\usepackage[colorinlistoftodos, textwidth=4cm, shadow]{todonotes}

\begin{document}

\title{Adaptive Online Learning with Momentum for Contingency-based Voltage Stability Assessment \\
}
\author{
\IEEEauthorblockN{
Zhijie Nie\IEEEauthorrefmark{1}\IEEEauthorrefmark{2},
Xiaohu Zhang\IEEEauthorrefmark{1},
Xiaoying Zhao\IEEEauthorrefmark{1},
Yiran Xu\IEEEauthorrefmark{3},
Di Shi\IEEEauthorrefmark{1}, 
Jiajun Duan\IEEEauthorrefmark{1} and
Zhiwei Wang\IEEEauthorrefmark{1}}
\IEEEauthorblockA{
\IEEEauthorrefmark{1}GEIRI North America, San Jose, CA 95134, USA\\
\IEEEauthorrefmark{2}Washington State University, Pullman, WA 99163, USA\\
\IEEEauthorrefmark{3}State Grid Nanjing Power Supply Company, Nanjing, China\\
Email: nie@ieee.org, xiaohu.zhang@geirina.net}

\thanks{This work was funded by SGCC Science and Technology Program.}
}



\maketitle


\begin{abstract}
Voltage stability refers to the ability of a power system to maintain acceptable voltages among all buses under normal operating conditions and after a disturbance. In this paper, a measurement-based voltage stability assessment (VSA) framework using online deep learning is developed. Since the topology changes induced by transmission contingencies may significantly reduce the voltage stability margin, different network topologies under different operating conditions are involved in our training dataset. To achieve high accuracy in the training process, a gradient-based adaptive learning algorithms is adopted. Numerical results based on the NETS-NYPS 68-bus system demonstrate the effectiveness of the proposed VSA approach. Moreover, with the proximal function modified adaptively, the adaptive algorithm with momentum outperforms traditional non-adaptive algorithms whose learning rate is constant.
\end{abstract}


\section{Introduction}

In recent years, with the advent of power system deregulation and increasing complexity of electricity consumption, the aging power grid has become congested and is under stress. The transmission facilities in today's competitive market are often operated close to their stability limits. Among various stability issues, voltage stability is one of the major concerns facing electric power utilities and regional transmission organizations (RTOs). To meet the system-wide voltage security requirement, it is crucial for the system operators to timely and accurately predict the voltage stability margin (VSM), i.e., the distance between current loading level and system loadability limit.

Traditionally, to accurately quantify the voltage stability margin (VSM), the continuation power flow (CPF) method is often used to address the relationship between load demand and the bus voltage magnitude by $PV$-curve. This method has been implemented in several commercial applications, such as PSS/E and DSATools. Alternatively, several types of voltage stability indices (VSIs) are derived as indicators to reveal the proximity to voltage collapse. Moghavvemi \textit{et al.} in~\cite{moghavvemi1998technique} propose the Line Stability Index (LSI) based on the feasibility in power flow solution through a transmission line. In \cite{musirin2002novel}, Fast Voltage Stability Index (FVSI) is presented to perform the voltage stability assessment (VSA). Based on the Thevenin equivalent circuit, the authors in \cite{xiaohu2018portland} propose a sensitivity-based index to track the VSM in real-time using local measurements. These aforementioned voltage stability indices can usually be computed very fast, which makes them suitable for online implementations.

The wide deployment of machine learning techniques provides an alternative approach addressing VSA with the collection of measurements. This category of VSA is also known as data-driven methods, which require no basic information of system network topology, the relationship between any system variables, or even the physical mechanism in power flow. It leverages system-wide measurements to train a model for particular usages. For instance, in~\cite{leonardi2011development}, Leonardi \textit{et al.} develop a regression model to seek a statistical relationship between the VSM and the reactive power reserves (RPR). Alternatively, in~\cite{li2017real}, Li \textit{et al.} utilize the convolutional neural network (CNN) to train a regression model between VSM and RPR. In~\cite{diao2009decision}, Diao \textit{et al.} develop a decision-tree-based classification model to predict the voltage stability of current operating condition under potential $N-1$ contingencies. Additionally, the authors propose the VSA framework, which consists of offline model training, periodic model updates, and online applications with PMU measurements. Instead of training the model from scratch, online learning is capable of resuming the training process by updating the predictor with new incoming data samples. In \cite{nie2017pmu}, an online learning framework is proposed to avoid re-trains on the ensemble of developed decision trees. This framework applies the \textit{Very Fast Decision Tree} (VFDT) induction algorithm to the online training. One advantage of this method lies in that it incorporates a pre-pruning step to prevent the overfitting issues. During the tree induction process, the node-splitting criterion becomes more conservative with the Hoeffding bound; however, the pre-pruning procedures only occur at the endings of the tree structures. Without changing the splitting criteria on the upper branches, the continuous growth on a decision tree will eventually give rise to overfitting issues when it encounters more and more updating data samples.



\textit{Artificial neural network} (ANN) is an alternative for online training. Reference \cite{malbasa2017voltage} proposes a voltage instability prediction method using ANN, which leverages gradient descent optimization algorithms. For the gradient descent method, a non-adaptive learning algorithm, each update is applied with a constant learning rate. This characteristic of non-adaptive learning algorithms places several challenges in finding the optimal weights of ANN models:

\begin{itemize}
    \item Selecting a good learning rate is usually difficult, which considerably depends on the the characteristics of dataset;
    \item Small values of learning rate may result in slow convergence, while large values may cause fluctuation around the optimal solution or may cause diverge within a few step;
    \item Schedules of decrements in learning rate is one solution to control the speed of convergence, but still requires experience on both tuning the rate and devising the schedules.
\end{itemize}

In light of these difficulties, this paper proposes an approach addressing VSA using adaptive online learning with momentum. To enhance the robustness in the training process, a large number of operating conditions (OCs) are generated based on different system topologies under different load levels. Subsequently, we apply $N-1$ contingency analysis to these OCs. The analysis outcomes indicate whether a configured OCs can withstand the contingencies without violating any operational limits. A multi-layer neural network model is trained, and the classification performance with adaptive learning in the gradient-based training algorithms is evaluated.

The remainder of this paper is organized as follows. Section \ref{sec::vsa} explains the concept of voltage stability, and describes the selection of transmission configurations based on the performance index of voltage. Section \ref{sec::gd_algos} demonstrates the variants of gradient-based algorithms used in online deep learning. In Section \ref{sec::case_study}, the numerical results based on the NETS-NYPS 68-bus test system are provided. Finally, the conclusions is presented in Section \ref{sec::conclu}.


\section{Voltage Stability Analysis considering Topology Changes} \label{sec::vsa}
\subsection{Vulnerability of System Operation}

\begin{figure}[!t]
    \centering
    \includegraphics[width=3.4in]{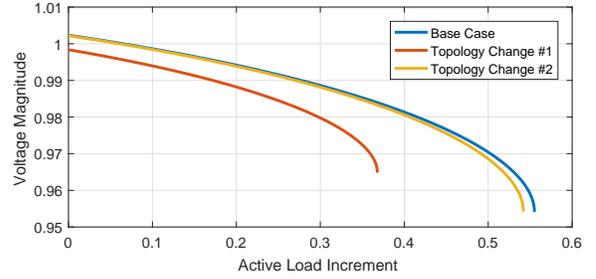}
    \caption{$PV$ curves subject to different topology changes.}
    \label{fig::pv_ctgs}
\end{figure}

Voltage Stability Analysis (VSA) evaluates the operating conditions to find system's maximum loadability to prevent voltage collapse. It is an indispensable tool for system operators to obtain timely situational awareness. In general, VSA evaluates the distance between the current operating point and the voltage collapse point, usually performed under a specific network topology. Nevertheless, in a practical system, topology change may occasionally occur due to instrumental failures, maintenance services, and some unexpected incidents. The impacts caused by topology changes on VSM are illustrated in \figurename~\ref{fig::pv_ctgs}. The base case illustrates the highest loadability limit with all branches in service, while under different topologies, the margins may vary: for topology change \#1, the system can withstand increasing loads until the voltage drops below 0.97 p.u.; however, in the case of topology change \#2, the maximum loadability maintains close to the base case. To improve the robustness in voltage stability prediction, it is essential to develop a measurement-based VSA that involves different network topologies in the training dataset.



To secure the system operation, contingency analysis is usually leveraged to evaluate whether there exists any violation when there is a line or generator outage in the system. Under the current OC, we conduct the contingency analysis from a list of contingencies. The result of contingency analysis indicates whether the OC can withstand these critical losses in the system. If the post-contingency case has a converged state without violating any limits, then the OC is labeled as ``Secure''; on the contrary, if the post-contingency case involves any operational violations or divergence, its OC is labeled as ``Insecure''. We develop the prediction model for measurement-based VSA as a two-class classifier. The training dataset of OCs is formed as $N$ pairs of $\left(x_i, y_i \right)$, in which $x_i \in \mathbb{R}^{M}$ is the OC vector consisting of $M$ measurements, and $y_i \in \{\text{Secure}, \text{Insecure}\}$ is the indicator from contingency analysis.

\subsection{Contingency Selection}

In this work, the performance index for voltage analysis described in \cite{mohamed1989voltage} is leveraged to select critical contingencies. Equation \eqref{eq::piv} demonstrates the voltage performance index for each system configuration ${c_k}$. This index evaluates the accumulated difference in voltage magnitudes between pre-configured OC and post-configured OC, i.e. $|V_i^{\text{pre}}|$ and $|V_i^{\text{post}}|$. ${\Delta |V_i|^{\text{lim}}}$ denotes the acceptable voltage deviation limit for each bus. $w_i$ denotes the weight considering the importance of a specific bus. And $n$ is the exponent for this index (typically, $n=1$ for the second-order $PI_{V}$).

\begin{equation}
    \label{eq::piv}
    PI_{V}^{c_k} = \sum_{i=1}^{N_\text{bus}} {\frac{w_i}{2n}}{\left(\frac{|V_i^{\text{post}}| - |V_i^{\text{pre}}|}{\Delta |V_i|^{\text{lim}}}\right)}^{2n}
\end{equation}

With the performance index calculated for each configuration, we can categorize the system configurations into two subclasses: one is \textit{topology change} (TC), which takes the scheduled outages into account, requiring additional assessments to ensure system operation security under $N-1$ contingencies; and the other is \textit{critical system contingency} (CSC), which causes distinct deviation in powerflow, and may lead to interruption of power supply during peak load periods. To reduce the space on system contingencies, we select the system configurations whose performance indices $PI_{V}$ are higher than 0.1 in this study. The system configurations with less than 200MW flow difference are considered as TCs, and the others are taken into account for VSA study as CSCs. 




\section{Gradient Descent Algorithms for Online Learning} \label{sec::gd_algos}

Gradient descent algorithms for online learning can be divided into two categories: the non-adaptive learning algorithms and the adaptive learning algorithms. 1) For \textit{non-adaptive learning} algorithms, the learning rate is set to be a constant for all weights' updates. 2) For \textit{adaptive learning} algorithms, the learning rate is adaptively modified according to the accumulated changes in each neuron's weight. A pre-defined schedule of learning rates is an intermediate solution to anneal the learning rate as the iteration proceeds applied in non-adaptive learning. However, such schemes require informative experience in the features and are unable to adapt to a dataset automatically \cite{ruder2016overview}. 




For simplicity, consider a basic unconstrained optimization setting to minimize the loss function. An iterative step to update the weights $\theta$ in the ANN model is 

\begin{equation}
    \theta_{k+1} = \theta_{k} + {\Delta {\theta}_{k}}
    \label{eq::param_update}
\end{equation}

In the following, we discuss different approaches to updating step ${\Delta {\theta}_{k}}$ in the gradient-based online learning. 

\subsection{Non-adaptive Gradient Decent Methods}
\subsubsection{Stochastic Gradient Descent (SGD)}
To prevent the training model from adapting to the ordering information, SGD shuffles the data at each epoch, then performs the update to each training sample. This method enables the online learning since each arriving sample has the potential to improve the trained model. For gradient-based algorithms, an update step is accomplished by the gradient descent direction in general:

\begin{equation}
    {\Delta {\theta}_{k}} = - \eta g_{k}
    \label{eq::sgd_next}
\end{equation}

\noindent
where $\eta > 0$ denotes the learning rate, and $g_{k} = \nabla_{\theta} f\left( \theta_{k} \right)$ denotes the gradient of loss function with respect to the current weight vector $\theta_{k}$. As such, gradient descent method assigns ${\Delta {\theta}_{k}}$ to follow the negative gradient direction of the loss function.

A momentum term $\gamma_k \in \left[0,1 \right]$ can be introduced to avoid oscillations by involving the most recent search direction ${\Delta {\theta}_{k-1}}$ in the iterative process as shown in \eqref{eq::sgdm_next}. 

\begin{equation}
    {\Delta {\theta}_{k}} = \gamma_k {\Delta {\theta}_{k-1}} - \eta g_{k}
    \label{eq::sgdm_next}
\end{equation}

\subsubsection{Nesterov Accelerated Gradient (NAG)}
With the momentum term introduced, we are approaching the optimal solution in a rapid trend; however, as the momentum accumulated, necessary responsiveness is critical to reduce the speed when the gradient direction starts to deflect. To achieve this, \cite{nesterov1983method} applies the gradient direction with respect to an approximation of the next position from the momentum term as follows.

\begin{equation}
    g_{k} = \nabla_{\theta} f\left( \theta_{k} + \gamma_k {\Delta {\theta}_{k-1}}  \right)
    \label{eq::nagm_grad}
\end{equation}

\noindent
Then we can apply \eqref{eq::sgdm_next} to update the neurons' weights. 

\subsection{Adaptive Gradient Descent Methods}\label{sec::ada_grad_desc}

\subsubsection{AdaGrad}
Overall, each update of weight in adaptive learning algorithms is computed by an individual learning rate. Based on the accumulative changes in weights, the adaptive learning approach performs smaller updates for weights that have been modified frequently, while it performs larger updates for the weights that only have mild changes in the previous iterations. AdaGrad in \cite{duchi2011adaptive} tunes the scales of individual updates according to the accumulation of previous subgradient for each weight:

\begin{equation}
    {\Delta {\theta}_{k}} = - \frac{\eta}{\sqrt{\sum_{i=1}^{k} {g_k^2}} + \epsilon} \circ g_{k}
    \label{eq::adag_next}
\end{equation}

\noindent
where the operator $\circ$ in \eqref{eq::adag_next} represents the elementwise multiplication of two vectors in the same dimension, and $g_{k}^2$ represents the elementwise square $g_{k} \circ g_{k}$. $\epsilon$ is a small value that avoids division by zero.

\subsubsection{Adaptive Moment Estimation (Adam)}
It can be seen that the learning rate of the weights within the initial steps decays rapidly in \eqref{eq::adag_next} with AdaGrad, which prevents the trained model from achieving higher accuracy. In \cite{kingma2014adam}, a more general version of AdaGrad called Adam is presented in order to handle these initialization bias. This approach includes bias-correction terms, and differentiates the updates for the gradients (first moments) and the squared gradients (second moments), respectively. It deploys the exponential moving averages to adaptively change the learning rates of weights:

\begin{equation}
    {\Delta {\theta}_{k}} = - \frac{\eta}{\sqrt{\hat{v}_k} + \epsilon} \circ \hat{m}_{k}
    \label{eq::adam_next}
\end{equation}

\noindent
where $\hat{m}_{k} = {m_k} / \left(1-\beta_{1}^{k}\right)$ and $\hat{v}_{k} = {v_k} / \left(1-\beta_{2}^{k}\right)$ are corrected estimates of the first raw moment ${m}_{k} = { \beta_{1} m_{k-1} + \left( 1 - \beta_{1} \right) g_{k} }$ and the second raw moment ${v}_{k} = { \beta_{2} v_{k-1} + \left( 1 - \beta_{2} \right) g_{k}^{2} }$ of the gradients respectively. $\beta_1$ and $\beta_2$ are the user-defined parameters that control the exponential decays in the learning rate, typically $\beta_1 = 0.9$, $\beta_2 = 0.999$.

With the Adam approach, the momentum methodology is implemented to involve the efforts in previous updates, and bias-correction terms introduced in the updates are able to avoid large initialization bias at the beginning of iterations.

\subsubsection{Nesterov-accelerated Adam (Nadam)}
Further, reference \cite{dozat2016incorporating} modifies NAG algorithm, and applies the update step as follows:

\begin{equation}
    {\Delta {\theta}_{k}} =  - \frac{\eta}{\sqrt{\hat{v}_k} + \epsilon} \circ \left( \beta_{1} \hat{m}_{k} + \frac{1 - \beta_{1}}{1-\beta_{1}^{k}} g_{k}  \right)
    \label{eq::nadam_next}
\end{equation}

For power system OCs, the studied case includes different types of measurements, e.g. voltage magnitudes, current measurements and the transferred active and reactive power on branches. Since different types of measurement data differ in scales and ranges, to achieve fast convergence and high accuracy, it is important to apply adaptive learning algorithms to the training process of the proposed measurement-based VSA.

\section{Case Study} \label{sec::case_study}

The NETS-NYPS 68-bus test system from Texas A\&M University Electric Grid Test Case Repository is selected to test the performance of the proposed VSA approach. The test system is partitioned into five areas with 68 buses, 16 generators, 64 transmission lines and 19 transformers. The total active and reactive power loads are 17620.7 MW and 2021.76 MVar respectively in the base case.

\begin{figure*}[!t]
    \centerline{\includegraphics[width=6.75in]{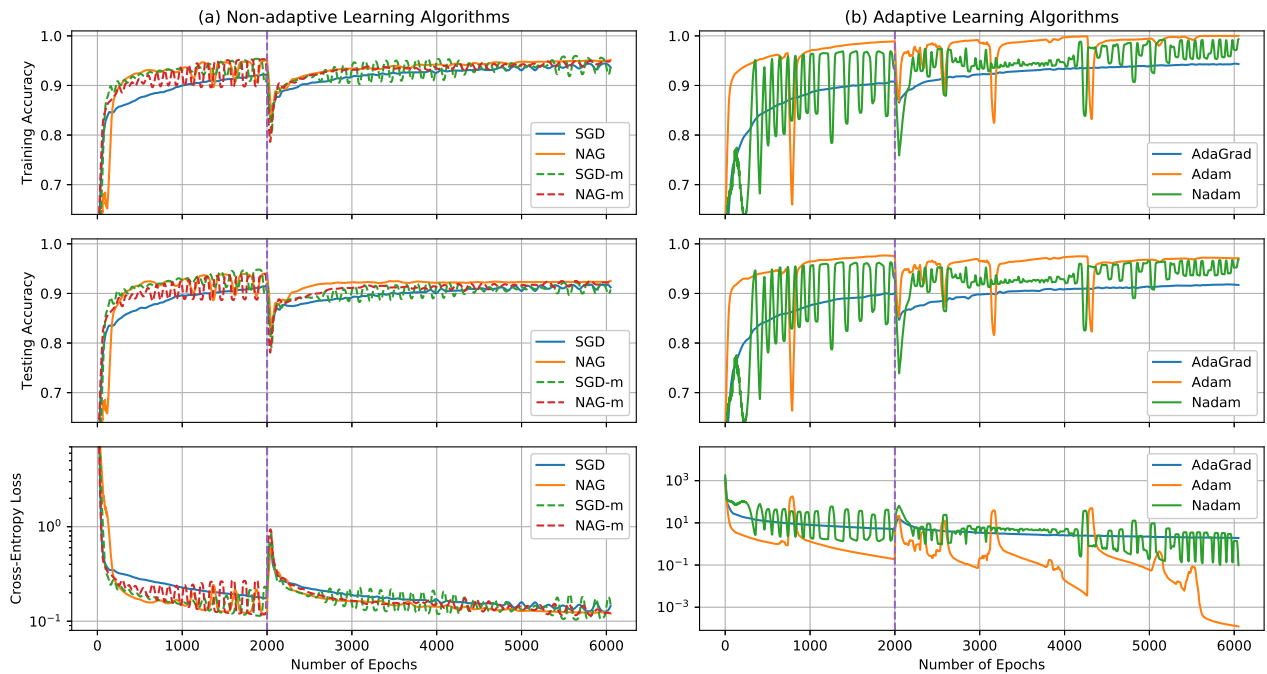}}
    \caption{Online training results for non-adaptive and adaptive learning algorithms.}
    \label{fig::combined_results}
\end{figure*}


\subsection{Preparation of Training Database of OCs}  \label{sec::oc_dataset}
The generation process of OCs is conducted by PSAT in DSATools. Each powerflow simulation considers the fluctuations in load demand, and each load is randomly scaled in a range $\left[0.8, 1.05 \right]$ from the base case. To balance the load deviations, the active power outputs are rescheduled based on their capacities. The simulation data collected for each OC includes voltage magnitudes/angles at load buses, current magnitudes and active/reactive power transferred on all the transmission lines. In total, there are 4,000 OCs generated in the dataset. Each OC consists of 524 features and is labeled with the contingency analysis results provided by VSAT. The dataset is divided into training and test sets with a 60-40 split.

\subsection{Deep Learning Framework}
To train the ANN model in an online learning manner, this study adopts the gradient descent algorithm to minimize the cross-entropy loss through backpropogations. To compare the performance in non-adaptive learning algorithms and adaptive learning algorithms, we study the algorithms demonstrated in Section \ref{sec::gd_algos}. The performances of these algorithms are compared in the number of learning epochs, cross-entropy loss, training accuracy and testing accuracy. In this work, a deep learning framework called MXNet~\cite{chen2015mxnet} is used for the multi-layer neural network training. The computation is accelerated using NVIDIA GeForce 940M GPU in Python 2.7 under Ubuntu 16.04 LTS environment.

\subsection{Training Performance}

\begin{table}[!t]
    \caption{Training accuracy achieved for continuous online training.}
    \begin{center}
        \begin{tabular}{c||cc|cccc}
            \hline
            \multirow{3}{*}{\textbf{Algorithm}} &  \multicolumn{6}{c}{\textbf{Number of Epochs}} \\
            \cline{2-7} 
            &  \multicolumn{2}{c|}{\textbf{\textit{Initialization Ph.}}} &  \multicolumn{4}{c}{\textbf{\textit{Update Ph.}}} \\
            \cline{2-7} 
            & \textbf{\textit{1000}}& \textbf{\textit{2000}} & \textbf{\textit{1000}} & \textbf{\textit{2000}} & \textbf{\textit{3000}} & \textbf{\textit{4000}} \\
            \hline
            SGD     & 0.8996 & 0.9229 & 0.9179 & 0.9261 & 0.9350 & 0.9454  \\
            NAG     & 0.9296 & 0.9489 & 0.9343 & 0.9439 & 0.9496 & 0.9507  \\
            SGD-m   & 0.9343 & 0.9489 & 0.9082 & 0.9329 & 0.9457 & 0.9557  \\
            NAG-m   & 0.9418 & 0.9532 & 0.9296 & 0.9421 & 0.9418 & 0.9518  \\ \hline
            AdaGrad & 0.8811 & 0.9079 & 0.9236 & 0.9307 & 0.9371 & 0.9418  \\
            Adam    & 0.9682 & 0.9893 & 0.9918 & 0.9971 & 0.9950 & 1.0000  \\
            Nadam   & 0.8357 & 0.9643 & 0.8650 & 0.8864 & 0.9900 & 0.9932  \\ \hline
        \end{tabular}
        \label{tab::acc_vs_epochs_train}
    \end{center}
\end{table}

\begin{table}[!t]
    \caption{Testing accuracy achieved for continuous online training.}
    \begin{center}
        \begin{tabular}{c||cc|cccc}
            \hline
            \multirow{3}{*}{\textbf{Algorithm}} &  \multicolumn{6}{c}{\textbf{Number of Epochs}} \\
            \cline{2-7} 
            &  \multicolumn{2}{c|}{\textbf{\textit{Initialization Ph.}}} &  \multicolumn{4}{c}{\textbf{\textit{Update Ph.}}} \\
            \cline{2-7} 
            & \textbf{\textit{1000}}& \textbf{\textit{2000}} & \textbf{\textit{1000}} & \textbf{\textit{2000}} & \textbf{\textit{3000}} & \textbf{\textit{4000}} \\
            \hline
            SGD     & 0.8967 & 0.9125 & 0.8780 & 0.8931 & 0.9006 & 0.9131  \\
            NAG     & 0.9175 & 0.9392 & 0.9215 & 0.9206 & 0.9206 & 0.9215  \\
            SGD-m   & 0.9250 & 0.9425 & 0.8931 & 0.9081 & 0.9190 & 0.9240  \\
            NAG-m   & 0.9300 & 0.9408 & 0.9123 & 0.9148 & 0.9173 & 0.9240  \\ \hline
            AdaGrad & 0.8742 & 0.9008 & 0.8989 & 0.9073 & 0.9173 & 0.9215  \\
            Adam    & 0.9642 & 0.9767 & 0.9724 & 0.9708 & 0.9683 & 0.9708  \\
            Nadam   & 0.8283 & 0.9642 & 0.8555 & 0.8747 & 0.9641 & 0.9691  \\ \hline
        \end{tabular}
    \label{tab::acc_vs_epochs_test}
    \end{center}
\end{table}

Generally, in machine learning, a model is trained under a specific dataset, which assumes the concept stationarity in the model. Online learning makes it possible to update the trained model when a new dataset is arrived. In this section, we present the online learning process as two phases. \textit{Initialization phase} indicates the learning process that the training model is built from scratch. \textit{Update phase} indicates the subsequent training process involving an update dataset so that the model can be trained to adapt to this new-incoming dataset.

To test the robustness of the training algorithms, the dataset in the initialization phase involves with no TC in the pre-configured OCs. In the update phase, we incorporate an update dataset considering multiple TCs in the pre-configured OCs as described in Section \ref{sec::oc_dataset}. The update dataset is combined with 70\% of normal OCs and 30\% OCs that involved with different individual TCs. Based on the performance index in Section \ref{sec::ada_grad_desc}, we consider eight individual TCs of the line outages at $l_{17-43}$, $l_{18-42}$, $l_{24-68}$, $l_{38-46}$, $l_{43-44}$, $l_{47-48}$, $l_{47-53}$, and $l_{54-55}$. The datasets in both phases consider the line outages at $l_{18-49}$, $l_{21-22}$, $l_{30-61}$, $l_{36-61}$, $l_{40-41}$, $l_{40-48}$, $l_{41-42}$, and $l_{67-68}$ as CSCs for contingency analysis. 



\figurename~\ref{fig::combined_results} illustrates the overall online learning process using non-adaptive learning and adaptive learning algorithms. ``-m'' denotes the algorithm that involves the momentum term $\gamma$ as shown in \eqref{eq::sgdm_next} and \eqref{eq::nagm_grad}. This work applies 2,000 epochs in the initialization phase and 4,000 epochs in the update phase. The halving vertical line indicates the switch between two phases. It can be observed that the update dataset causes the sudden drop in the training/testing accuracy between two phases. For non-adaptive learning algorithms in the initialization phase, NAG, SGD-m, and NAG-m are able to achieve better prediction accuracy than SGD due to the modifications in the update steps illustrated in Section  \ref{sec::gd_algos}. It can be also observed that with momentum terms involved, both SGD-m and NAG-m algorithms are capable of achieving higher accuracy than SGD and NAG training within a very small number of epochs. 

Table \ref{tab::acc_vs_epochs_train} and Table \ref{tab::acc_vs_epochs_test} show training/testing accuracy at the training steps with 1,000, 2,000 epochs and at the continued training steps from 1,000 to 4,000 epochs. From the two tables we observe that, even after the learning process is incorporated with a new dataset in the update phase, the ANN-based online training methods can still achieve high-accuracy performance as it previously achieved. Both Adam and Nadam show their significant competence in loss minimization, and the early-stopping criteria could be devised when the error is within a satisfactory range.

\section{Conclusions} \label{sec::conclu}

In this study, we focus on online deep learning methods for the measurement-based VSA. To improve the situational awareness under different topologies, several topology changes are involved in the pre-configured OCs, then the $N-1$ contingency analysis is conducted by VSAT. Consequently, the trained predictor is able to provide the post-contingency security status based on the current measurements. To demonstrate the training performance in different online learning methods, this paper compares AdaGrad, Adam, and Nadam and the training results show adaptive learning algorithms can achieve higher accuracy even when the online learning process is incorporated with a new dataset at the update phase. In the training of measurement-based VSA predictor, Adam outperforms the other two algorithms in training/testing accuracy, loss minimization, and algorithm stability. Future research will continue on the study of online system security prediction with a finite number of measurement collections since the real-time measurements are not always available, and only a limited number of voltage phasors can be captured by synchrophasors.



\bibliographystyle{IEEEtran}
\bibliography{ref}

\begin{thebibliography}{10}
\providecommand{\url}[1]{#1}
\csname url@samestyle\endcsname
\providecommand{\newblock}{\relax}
\providecommand{\bibinfo}[2]{#2}
\providecommand{\BIBentrySTDinterwordspacing}{\spaceskip=0pt\relax}
\providecommand{\BIBentryALTinterwordstretchfactor}{4}
\providecommand{\BIBentryALTinterwordspacing}{\spaceskip=\fontdimen2\font plus
\BIBentryALTinterwordstretchfactor\fontdimen3\font minus
  \fontdimen4\font\relax}
\providecommand{\BIBforeignlanguage}[2]{{%
\expandafter\ifx\csname l@#1\endcsname\relax
\typeout{** WARNING: IEEEtran.bst: No hyphenation pattern has been}%
\typeout{** loaded for the language `#1'. Using the pattern for}%
\typeout{** the default language instead.}%
\else
\language=\csname l@#1\endcsname
\fi
#2}}
\providecommand{\BIBdecl}{\relax}
\BIBdecl

\bibitem{moghavvemi1998technique}
M.~Moghavvemi and F.~Omar, ``Technique for contingency monitoring and voltage
  collapse prediction,'' \emph{IEE Proceedings-Generation, Transmission and
  Distribution}, vol. 145, no.~6, pp. 634--640, 1998.

\bibitem{musirin2002novel}
I.~Musirin and T.~A. Rahman, ``{Novel fast voltage stability index (FVSI) for
  voltage stability analysis in power transmission system},'' in \emph{Research
  and Development, 2002. SCOReD 2002. Student Conference on}.\hskip 1em plus
  0.5em minus 0.4em\relax IEEE, 2002, pp. 265--268.

\bibitem{xiaohu2018portland}
X.~Zhang, D.~Shi, X.~Lu, Z.~Yi, Q.~Zhang, and Z.~Wang, ``Sensitivity based
  thevenin index for voltage stability assessment considering n-1
  contingency,'' in \emph{2018 IEEE Power Energy Society General Meeting},
  August 2018, pp. 1--5.

\bibitem{leonardi2011development}
B.~Leonardi and V.~Ajjarapu, ``Development of multilinear regression models for
  online voltage stability margin estimation,'' \emph{IEEE Transactions on
  Power Systems}, vol.~26, no.~1, pp. 374--383, 2011.

\bibitem{li2017real}
S.~Li and V.~Ajjarapu, ``Real-time monitoring of long-term voltage stability
  via convolutional neural network,'' in \emph{2017 IEEE Power Energy Society
  General Meeting}, July 2017, pp. 1--5.

\bibitem{diao2009decision}
R.~Diao, K.~Sun, V.~Vittal, R.~J. O'Keefe, M.~R. Richardson, N.~Bhatt,
  D.~Stradford, and S.~K. Sarawgi, ``Decision tree-based online voltage
  security assessment using pmu measurements,'' \emph{IEEE Transactions on
  Power Systems}, vol.~24, no.~2, pp. 832--839, 2009.

\bibitem{nie2017pmu}
Z.~Nie, D.~Yang, V.~Centeno, and K.~D. Jones, ``A pmu-based voltage security
  assessment framework using hoeffding-tree-based learning,'' in \emph{2017
  19th International Conference on Intelligent System Application to Power
  Systems (ISAP)}, Sept 2017, pp. 1--6.

\bibitem{malbasa2017voltage}
V.~Malbasa, C.~Zheng, P.-C. Chen, T.~Popovic, and M.~Kezunovic, ``Voltage
  stability prediction using active machine learning,'' \emph{IEEE Transactions
  on Smart Grid}, vol.~8, no.~6, pp. 3117--3124, 2017.

\bibitem{mohamed1989voltage}
A.~Mohamed and G.~Jasmon, ``Voltage contingency selection technique for
  security assessment,'' in \emph{IEE Proceedings C (Generation, Transmission
  and Distribution)}, vol. 136, no.~1.\hskip 1em plus 0.5em minus 0.4em\relax
  IET, 1989, pp. 24--28.

\bibitem{ruder2016overview}
S.~Ruder, ``An overview of gradient descent optimization algorithms,''
  \emph{arXiv preprint arXiv:1609.04747}, 2016.

\bibitem{nesterov1983method}
Y.~Nesterov, ``{A method for unconstrained convex minimization problem with the
  rate of convergence O(1/k\^{}2)},'' in \emph{Doklady AN USSR}, vol. 269,
  1983, pp. 543--547.

\bibitem{duchi2011adaptive}
J.~Duchi, E.~Hazan, and Y.~Singer, ``Adaptive subgradient methods for online
  learning and stochastic optimization,'' \emph{Journal of Machine Learning
  Research}, vol.~12, no. Jul, pp. 2121--2159, 2011.

\bibitem{kingma2014adam}
D.~P. Kingma and J.~Ba, ``Adam: A method for stochastic optimization,''
  \emph{arXiv preprint arXiv:1412.6980}, 2014.

\bibitem{dozat2016incorporating}
T.~Dozat, ``Incorporating nesterov momentum into adam,'' \emph{ICLR Workshop},
  2016.

\bibitem{chen2015mxnet}
T.~Chen, M.~Li, Y.~Li, M.~Lin, N.~Wang, M.~Wang, T.~Xiao, B.~Xu, C.~Zhang, and
  Z.~Zhang, ``Mxnet: A flexible and efficient machine learning library for
  heterogeneous distributed systems,'' \emph{Neural Information Processing
  Systems, Workshop on Machine Learning Systems}, 2016.

\end{thebibliography}



\end{document}